\documentclass{kapproc} 

\usepackage{graphicx}

\kluwerbib
\let\lcitebracket(
\let\rcitebracket)

\newcommand{\msun}{\mbox{${\rm M}_{\odot}$}}

\newcommand{\mc}{\mbox {$M_{Ch}$}}
\newcommand{\ms}{\mbox {${\rm M}_{\odot}$}}
\newcommand{\rs}{\mbox {$R_{\odot}$}}
\newcommand{\myr}{\mbox {${\rm M_{\odot}~yr^{-1}}$}}

\def\apgt{\ {\raise-.5ex\hbox{$\buildrel>\over\sim$}}\ }
\def\aplt{\ {\raise-.5ex\hbox{$\buildrel<\over\sim$}}\ }

\let\footnote\savefootnote
\let\footnoterule\savefootnoterule 

\begin{document}

\articletitle[Close Double White Dwarfs]{The Population of Close Double White Dwarfs in the Galaxy}

\author{Lev R. Yungelson}
\affil{Institute of Astronomy of the Russian Academy of
             Sciences,
 Moscow, Russia
}
\author{Gijs Nelemans}
\affil{Astronomical Institute ``Anton Pannekoek'', 
Amsterdam, the Netherlands}
\author{Simon F.
  Portegies Zwart\thanks{Hubble Fellow}}
\affil{Massachusetts Institute of Technology,
 Cambridge, USA}
\author{Frank  Verbunt}
\affil{Astronomical Institute, Utrecht University,
Utrecht, the Netherlands}


\begin{abstract} 
We present a new model for the Galactic population
of close double white dwarfs. The model accounts for the suggestion
of the avoidance of a substantial spiral-in during mass transfer between a giant and a main-sequence star  of comparable mass
and for detailed cooling models. It agrees  well with the 
observations of the local sample of white dwarfs  if the initial
binary fraction is $\sim\!50$\,\% and an {\it ad hoc} assumption is
made that white dwarfs with $M\!\aplt 0.3\,\ms$ cool faster than the
models suggest. About 1000 white
dwarfs  with $V\!\aplt\!15$ have to be surveyed for detection of a pair which  has $M_1+M_2\!\apgt\!\mc$ and will merge within
10\,Gyr. 
\end{abstract}

\section{Introduction}\label{intro}
The interest in the close double white dwarfs (hereafter CDWD) stems from several reasons: (i) white dwarfs are the endpoints of stellar evolution; (ii)  CDWD experienced at least two stages of mass exchange and thus provide an important tool for testing the evolution of binaries; 
(iii) merging double CO white dwarfs are a model for 
SNe\,Ia;
(iv) CDWD may be the most important contributors to the gravitational
wave noise  at $\nu \aplt 0.01\,{\rm Hz}$, possibly burying  signals of other sources. 

We review the data on the currently known CDWD and present results of the modelling of the population of white dwarfs, which involves several new aspects.
The most important are the treatment of the first stage of mass loss without significant 
spiral-in \cite{nvy+00}, the use of detailed models for the cooling of
white dwarfs  
and the consideration of different star formation histories for the Galactic disc. 

\section{Observed close double white dwarfs}\label{sec:obs}
\begin{table}[!t]
\caption[]{Known close double white dwarfs 
and subdwarfs  with WD companions}
\begin{tabular*}{\textwidth}{@{\extracolsep{\fill}}rllcc} 
\sphline
{\it N} & {\it Name} & $P (d)$ & $M_1/M_\odot$  & $M_2/M_\odot$  \\ 
\sphline
1 & WD~0135$-$052 & 1.556  & 0.47 & 0.52\\
2 & WD~0136$+$768 & 1.407  & 0.44 & 0.34\\
3 & WD~0957$-$666 & 0.061  & 0.37 & 0.32\\
4 & WD~1101$+$364 & 0.145  & 0.31 & 0.36\\ 
5 & WD~1204$+$450 & 1.603  & 0.51 & 0.51\\ 
6 & WD~1704$+$481A & 0.145  & 0.39 & 0.56\\ 
7 & WD~1022$+$050 & 1.157  & 0.35 &     \\
8 & WD~1202$+$608 & 1.493  & 0.40 &     \\
9 & WD~1241$-$010 & 3.347  & 0.31 &     \\
10 & WD~1317$+$453 & 4.872  & 0.33 &     \\
11 & WD~1713$+$332 & 1.123  & 0.38 &     \\
12 & WD~1824$+$040 & 6.266  & 0.39 &     \\
13 & WD~2032$+$188 & 5.084  & 0.36 &     \\
14 & WD~2331$+$290 & 0.167  & 0.39 &     \\ 
\sphline
15 & KPD~0422+5421 & 0.09 & 0.51 & 0.53  \\
16 & KPD~1930+2752 & 0.095 & 0.5 &  0.97 \\ 
\sphline
\end{tabular*}
\begin{tablenotes}
Objects 1 - 14 are CDWD, objects 15 and 16 are sdB stars with suspected white dwarf companions.  $M_1$ is the mass of the brighter component and    $M_2$ is the mass of the companion. See the text for references.
\end{tablenotes}
\label{tab:obs}
\end{table}

Table~\ref{tab:obs} lists the currently known CDWD for
which the orbital period and the mass of at least one component are
measured \cite{mm99,mmmh00}. The accuracy of the mass determinations is
certainly not better than $\pm 0.05\,\ms$. With the exception of
WD~0135$-$052 and 1204$+$450, the brighter components of the pairs are
likely to be He white dwarfs, since their mass is lower than
0.46\,\ms, the limiting mass for the helium ignition in a degenerate stellar core
\cite{sgr90}. In the range of  $M\simeq\!(0.35 - 0.45)\,\ms$\ white
dwarfs could have CO cores and thick He envelopes \cite{it85}.
However, the probability of the formation of these so called ``hybrid''
WD is 4 -- 5 times lower than for  helium WD with the same mass
\cite{ty93,nyp+00}.  The binary dwarfs 
WD~0957$-$666, 1101$+$364, 1704$-$481A, and 2331$+$290 are expected to merge
because of  angular momentum loss via gravitational wave radiation.
If both components are He dwarfs, the merger may result in
the formation of a nondegenerate star or a supernova-scale
explosion \cite{ns77}.  In the case of CO companions, the formation of an R\,CrB star is expected \cite{web84,ity96}.

We included in Tab.~\ref{tab:obs} also data on two sdB stars  with
white dwarf companions \cite{ow99,mmn00}. 
Subdwarfs are supposed to be helium-burning stars. In these particular systems central helium burning will be completed before  components merge. This makes these binaries candidates for future merging CO+CO white dwarfs. Remarkably, for KPD~1930$+$2752 the total mass of the system exceeds \mc, making it a possible SN\,Ia candidate! 

About 10 more WD and sdB stars with suspected close WD companions are known  \cite{mar00,mmm001}. However, for these systems the orbital periods or the masses of the components are not yet determined.

\section{Formation of helium white dwarfs}
\label{sec:form}

Three CDWD are of special interest -- WD~0136$+$768, 0957$-$666, and 1101$+$364. In these systems  the masses of both components suggest that they are helium dwarfs and, thus, descend from degenerate cores of low-mass (sub)giants.
If we designate main-sequence stars as  MS, red (sub)giants as  RG,   white dwarfs as WD, the stage of mass exchange, either stable or unstable, as RLOF, and refer by subscripts 1 and 2 to the initial primary and secondary star, the evolutionary sequence which results in the formation  of a double helium WD may be described as follows:
MS$_1$ + MS$_2$ $\rightarrow$ RG$_1$+MS$_2$ $\rightarrow$ RLOF$_1$ $\rightarrow$ WD$_1$+MS$_2$   $\rightarrow$WD$_1$+RG$_2$
$\rightarrow$ RLOF$_2$ $\rightarrow$ WD$_1$+WD$_2$.
If the mass transfer is unstable, the change of component separation $a$ is usually calculated by balancing the binding energy of the envelope of the mass-losing star with the change of the orbital energy \cite{web84}:  
\begin{equation}\label{eq:ce}
\frac{M_1 \; (M_1- M_c)}{\lambda \; r_{\rm 1}}  
 = \alpha \; \left[ \frac{M_c\; M_2}{2 \;a_{\rm
       f}} -  \frac{M_1\; M_2}{2
 \;a_{\rm i}} \right].
\end{equation}
Here $M_c$ is the mass of the core of the mass-losing star, $r_1$ is its radius,  subscripts $i$ and $f$ refer to the initial and final values of the orbital separation, $\alpha$\ is the efficiency of the deposition of orbital energy into the common envelope and $\lambda$ is a parameter which depends on the density distribution in the stellar envelope; the  usual assumption is $\lambda=0.5$~ \cite{khp87}. It has to be noticed,  however,  that for stars more massive than 3\,\ms\ the values of $\lambda$\ for red giants may be significantly lower \cite{dt00}. It would be worthwhile to investigate also  lower mass stars, which form most of the observed CDWD. 

Giants with degenerate helium cores obey a  
unique core-mass -- radius relation \cite{rw70}. 
Neglecting a slight dependence on the total mass of the star,  this dependence (for solar chemical composition objects) may expressed as  
\cite{it85}:
\begin{equation}\label{eq:Rg}
R  \approx 10^{3.5}  \; M_{\rm c}^4,
\end{equation}
where radius of the giant $R$ and mass of the core $M_{\rm c}$ are in solar units. At the instant when the star fills its Roche lobe, the radius given by Eq.~(\ref{eq:Rg}) is equal to the radius of the Roche lobe.   If the mass loss is unstable, the mass of the core of the mass-losing star doesn't grow during this stage and the mass of the  new-born white dwarf is equal to $M_{\rm c}$.

\begin{figure*}[!ht] 
\begin{tabular}{lr}
\begin{minipage}[t]{0.48\textwidth}
\includegraphics[angle=-90,scale=0.255]{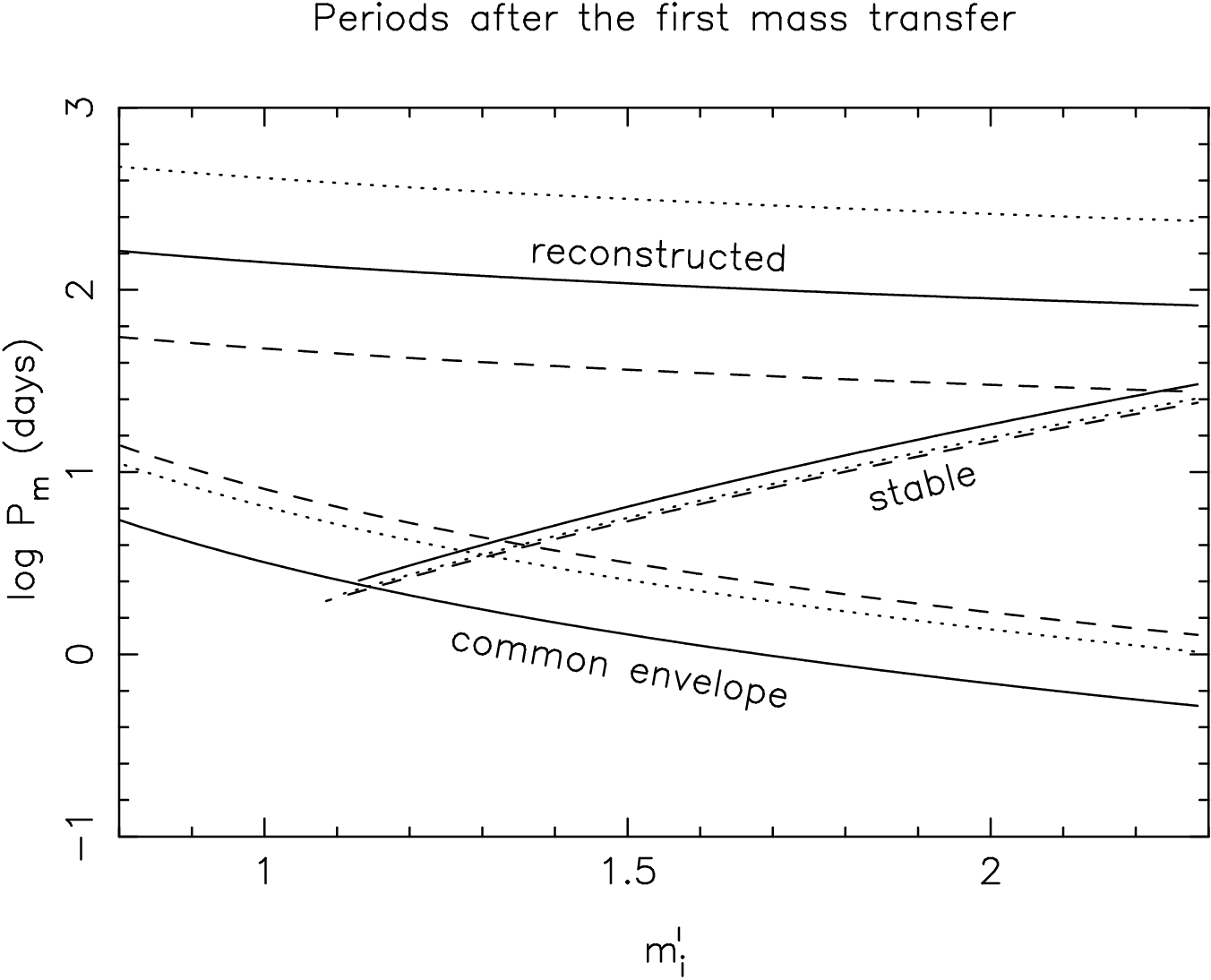}
\caption[]{Periods after the first phase of mass loss as a function
of the mass of the secondary component at this time. Solid, dashed,
and dotted  lines are for WD 0957-666, 1101+364, and 0136+768
respectively. The top three lines are periods needed to explain the
mass of the last formed white dwarf. The middle  three lines give the
maximum period  if the first white dwarf would be formed by
conservative mass exchange. The lower three lines give the periods
for the case when the  formation of the dwarf  was accompanied by a
spiral-in (Eq.~(\ref{eq:ce}) with $\alpha\lambda =2$). }
\label{fig:Pm_log}
\end{minipage}
\hspace{0.02\textwidth}
\begin{minipage}[t]{0.48\textwidth}
\includegraphics[angle=-90,scale=0.275]{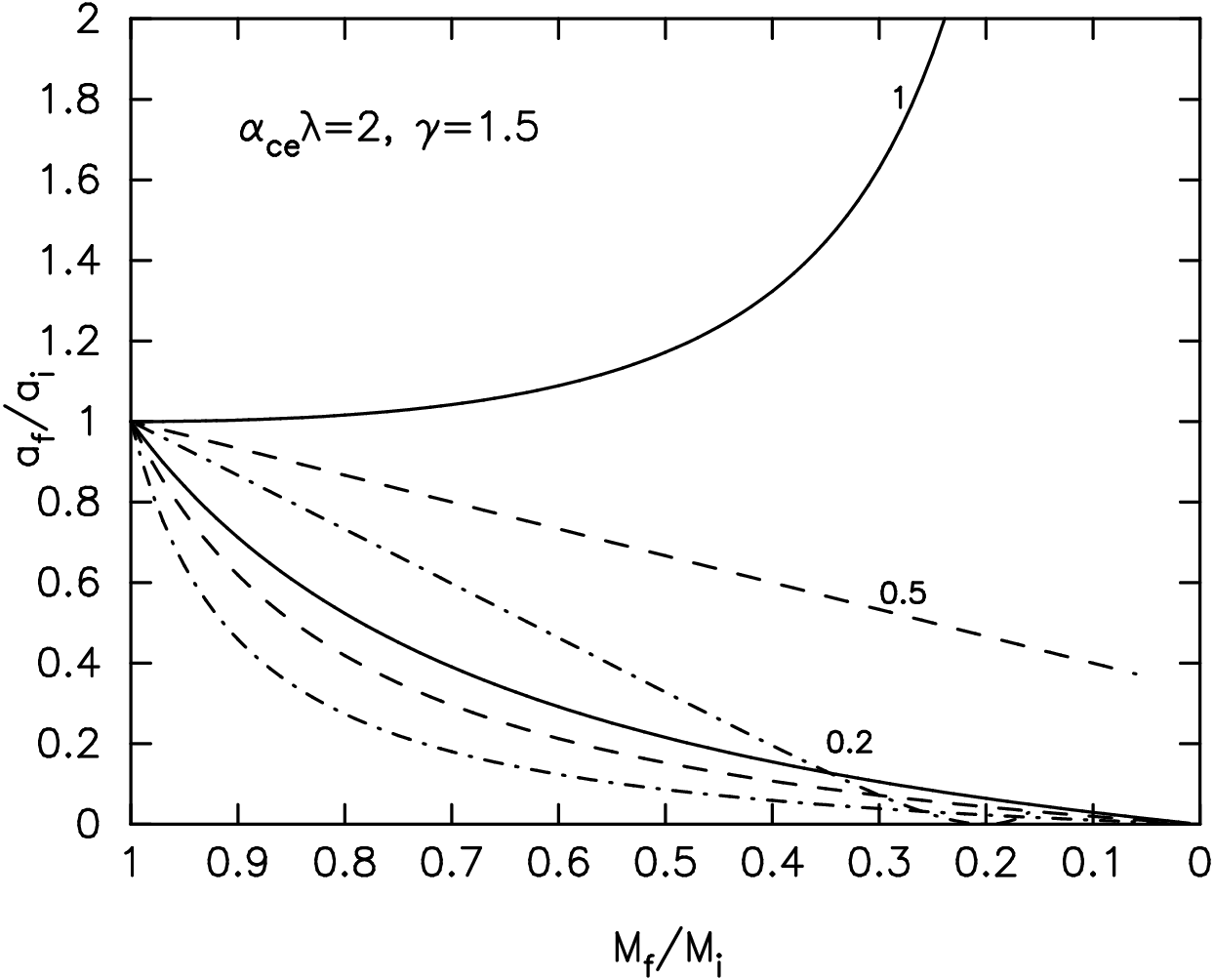}
\vspace{0.5mm} 
\caption[]{Dependence of the relative
variation of the separation of the components $a_f/a_i$ on the
fraction of the mass lost by the star for the cases of spiral-in in a
common envelope and AML regulated variation of $a$. Pairs of solid,
dashed, and dash-dotted curves correspond to the initial mass ratio
of 1.0, 0.5, and 0.2, respectively. The upper curve of every pair is
for the ``AML formalism'' [Eq.~(\ref{eq:aml})]. } \label{fig:afai15}
\end{minipage} 
\end{tabular} 
\end{figure*}

Applying Eq.~(\ref{eq:Rg}) together with restrictions on the masses of the progenitors of helium white dwarfs (0.8 -- 2.3 \ms), Nelemans et al. (2000a)\nocite{nvy+00} reconstructed the evolution of WD~0136$+$768, 0957$-$666, and 1101$+$364. 
Their conclusions  may be summarized as follows: (i)  in the first episode of mass loss, when the companion of the Roche lobe filling star was  a main-sequence star of a comparable mass, no substantial spiral-in occurred, see Fig.~\ref{fig:Pm_log}; (ii) in the second mass loss episode, which resulted in the formation of the currently brighter white dwarf, the separation of the components was strongly reduced.  The deposition of the energy into common envelope in this episode was highly efficient:  in Eq.~(\ref{eq:ce}) the product $\alpha\lambda \aplt 3$. A note has to be made in relation with the latter statement: since Eq.~(\ref{eq:ce}) provides nothing more than an order of magnitude estimate,  a formal solution which gives $\alpha \geq 1$\   
indicates that energy deposited into common envelope has to be comparable to the orbital energy of the binary.  

Since the first mass loss episode is neither  stable mass transfer nor a spiral-in in a common envelope and the physical picture of the process  is absent, Nelemans et al. (2000a) \nocite{nvy+00} suggested to use in the population synthesis studies a simple {\em linear} equation for the angular momentum balance: 
\begin{equation}\label{eq:aml}
J_{\rm i} - J_{\rm f} = \gamma J_{\rm i} \frac{\Delta M}{M},
\end{equation}
where $J_{\rm i}$ and $J_{\rm f}$ are  the angular momenta of the pre- and post-mass-transfer
binary, respectively, $\Delta M$\ is the amount of mass lost by the binary and $M$ is the total initial mass of the system. The parameter $\gamma$ is adjusted by fitting the orbital periods and masses of the three abovementioned helium CDWD. It follows that
$1.4\!\aplt\!\gamma\!\aplt\!1.8$. As Fig.~\ref{fig:afai15} illustrates, when Eq.~(\ref{eq:aml}) is applied, the separation between the components changes much less drastically, compared to the case when the ``standard'' common envelope formalism [Eq.~(\ref{eq:ce})] is applied. For typical values of the fractional mass of the core $M_f/M_i$\  both formalisms give similar $a_f/a_i$ when  $q$ decreases to $\simeq 0.2$. In the actual calculations, for a given $M_f/M_i$, the larger of two values of $a_f/a_i$ was used.

\section{Cooling of white dwarfs and observational selection}\label{sec:cool}

\begin{figure}[!th]
\begin{minipage}[!h]{0.48\textwidth}
\hskip -4mm
\includegraphics[angle=-90,scale=0.28]{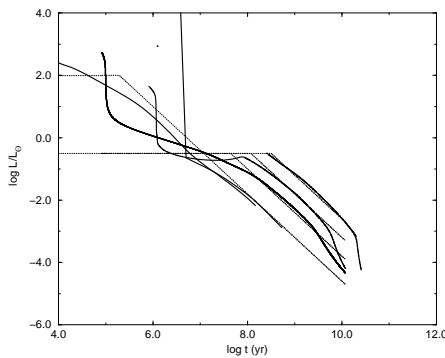}
\end{minipage}
\begin{minipage}[!th]{0.48\textwidth}
\caption[]{Cooling curves  for 0.179, 0.300, 0.414\,\ms\ white dwarfs from Driebe et al. (1998)\nocite{dsb+98}  and for 0.6 and 0.8\,\ms\ ones from 
Bl\"ocker (1995)\nocite{blo95b}, from  right to left. Straight lines are the fits to the curves used in the simulations.   
}
\end{minipage}
\label{fig:wdcool}
\end{figure}

The observed sample of CDWD is biased. Some objects were selected for study since their low mass already suggested binarity.  White dwarfs must be sufficiently bright for the mass determination and the measurement of the radial
velocities. This suggests to compare  a magnitude limited model sample of dwarfs with the observations. Hitherto, following Iben and Tutukov (1985)\nocite{it85}, it was assumed in all population synthesis studies  that all WD can be observed for $10^8$\,yr, unless close pairs merge in a shorter time. The actual cooling curves were never applied. However, recent studies \cite{dsb+98,han99,ba99,sea00}  show, that helium WD cool more slowly than carbon-oxygen ones. This is due to the higher heat content of the helium WD \cite{han99} and  residual nuclear burning in the relatively thick hydrogen envelopes (first noticed by Webbink (1975)\nocite{web75}). In our basic model, we take cooling curves for CO dwarfs from 
Bl\"ocker (1995)\nocite{blo95b} and for He ones from Driebe et al. 
(1998, hereafter DSBH)\nocite{dsb+98}, see Fig.~\ref{fig:wdcool}. The initial models for these curves  are obtained by mimicking the mass loss by stars and therefore may be considered as the most realistic. However, as we show below, cooling times given by them  probably can not be taken at  face value and need revision downward. Mass loss in stellar wind and thermal flashes, which extinguishes hydrogen burning \cite{it86b,sea00,asb00}, may provide the necessary mechanism. 

In addition to the sufficient brightness of the components, CDWD  must have such orbital periods  that  
radial velocities would be large enough to be measured, but small enough not to be smeared
out during the integration.   Following estimates of Maxted and Marsh (1999)\nocite{mm99}, we model this selection effect assuming that CDWD with $0.15\,{\rm hr} \leq P_{orb} \leq  8.5\,{\rm day}$\ will be detected
with 100\% probability and that above 8.5\,day the detection
probability decreases linearly  to 0 at $P_{orb}=35\,
{\rm day}$.

Our other basic assumptions may be listed as follows: we use Miller and Scalo (1979)\nocite{ms79} IMF, flat distributions over initial mass ratios of the components and the logarithm of the orbital separation, and a thermal distribution of orbital eccentricities. 
	 
\section{An overview of the models}\label{sec:res}

For the modelling of the population of double white dwarfs we used the numerical code \textsf{SeBa} \cite{pv96} with modifications for low- and intermediate mass stars described by Nelemans et al. (2000b)\nocite{nyp+00}. Some of the important points are discussed below. 

Since most progenitors of white dwarfs  are low-mass stars, the
Galactic star formation history  influences the  current birthrate of
CDWD  and the properties of their population.  This factor  was not
studied before. For the present study  we take a time-dependent  star
formation rate in the Galactic disk as 
\begin{equation}\label{eq:sfr} {\rm SFR}(t) = 15 \; \exp(-t/\tau)
\quad \ms \; \mbox{yr}^{-1}, \end{equation} 
where $\tau$ = 7 Gyr.
Equation~(\ref{eq:sfr}) gives current SFR of 3.6\,\myr, which is
compatible with the observational estimates \cite{ran91,hj97}. With
this equation,  the amount of matter that has been turned into stars
over the lifetime  of the Galactic disk (10 Gyr in our model) is
$\sim 8 \times 10^{10}$\,\msun.  It is higher than the current mass
of the disk, since a part of this matter is returned  to the ISM by
supernovae and stellar winds.  We also compute several models with
constant SFR, to allow comparison with previous work (see
Tab.~\ref{tab:mod}).  

The distribution of  stars in the Galactic disk  is taken as 
\begin{equation}\label{eq:rho}
\rho(R, z) = \rho_{\rm 0} \; e^{-R/H} \; \mbox{{\rm sech}}(z/h)^2  \quad \mbox{pc}^{-3},
\end{equation}
where we use $H$ = 2.5 kpc \cite{sac97} and $h$ = 200 pc.
The age and mass dependence of $h$ is neglected.

Table~\ref{tab:mod} gives an overview of our assumptions and model results and a comparison with some models of other authors. Model A is our basic model with an exponential star formation rate in the disk [Eq.~(\ref{eq:sfr})], initial fraction of binaries equal to 50\% (i.e. with 2/3 of all stars in binaries)  and cooling of white dwarfs according to DSBH, but modified as described below, in Sec.~\ref{sec:mod_vs_obs}. The 50\%\ fraction of binaries is suggested as a lower limit to their actual occurrence  by  the studies of normal main-sequence stars \cite{abt83,dm91}. This model, as well as all our models presented in the Table, were calculated with $\alpha\lambda=2$\ in Eq.~(\ref{eq:ce}) and $\gamma=1.75$\ in Eq.~(\ref{eq:aml}). Model  A$^\prime$ is similar to the model A, but assumes that the first stage of mass loss  is a common envelope described by Eq.~(\ref{eq:ce}) instead of Eq.~(\ref{eq:aml}). 
Model B is similar to the model A, but has initially all stars in  binaries, while model C has a constant star formation rate and 50\%  binaries. Model D has a constant star formation rate and 100\%  binaries. Models C and D were normalised  to SFR  of 4\,\myr, to allow comparison with the models of Iben et al. (1997, ITY)\nocite{ity97} and Han (1998)\nocite{han98}. The former model was  recalculated by LRY for the disk age of 10\,Gyr, the age assumed in this study. Note, that ITY assume that  unstable mass loss always results in the formation of a common envelope, and their formulation of the equation for the energy balance is different from  Eq.~(\ref{eq:ce}).    

\begin{table}[!t]
\caption[]{Summary of models}
\begin{tabular*}{\textwidth}{@{\extracolsep{\fill}}lrrcccccccc}
\sphline
{\it Mod.}   & {\it SFR} &  {\it bin}  & {\it WD} & $\nu$  & $\nu_{\rm m}$ &
  {\it SN\,Ia} & {\it IWD} & {\it CWD} &  $\rho_{wd,\odot}$ & $\nu_{\rm PN,\odot}$ \\ 
  & & \% & $10^9$ & $10^{-2}$ & $10^{-2}$ &  $10^{-3}$ & $10^{-3}$ & $10^8$ &  $10^{-3}$ & $10^{-12}$\\
\sphline
  A       & Exp & 50  &  9.2 & 4.8 & 2.2 & 3.2 & 4.6 & 2.5 &  19 & 2.3 \\    A$^\prime$ & Exp & 50  & & 3.0  & 2.4 & 2.2 & 11.0 & 1.0  & & \\
  B       &Exp     & 100 & & 8.1 &  3.6 & 5.4 & 8.8 & 4.1 & & \\ 
  C       &Cns     & 50  & 4.0 & 3.2 &  1.6 & 3.4  & 3.6 & 1.2 &  8.5 & 1.9 \\
  D       &Cns& 100 & & 5.3  & 2.8 &  5.8 & 6.1 & 1.9 &   & \\[0.1cm] 
  HAN   &Cns& 100& & 2.9 & 2.8  & 2.6 & 23.0& 0.9 & &  \\
  ITY   &Cns &100& 4.0 &  8.7 & 1.7 & 1.9 & 8.5 & 3.5 & 8.3  & 1.5 \\ [0.1cm]
 
  OBS     &    &    &      &  &      & $4\pm2$&   &     & $4\div20$ & 3 \\  \sphline
\end{tabular*}
\begin{tablenotes}
The columns list the identifiers of the models, type of star formation rate assumed for the model (exponential or constant), initial fraction of binaries, total Galactic number of WD, rates  of formation and  merger of CDWD per 100\,yr, rate of merger of CDWD with $M_1+M_2 \geq \mc$ per 1000\,yr (SNe\,Ia), rate of formation of Interacting double WD (AM\,CVn type stars) per 1000\,yr, total number of Close double WD in the Galaxy, local density of all  WD per ${\rm pc^3}$, local rate of formation of planetary nebulae per ${\rm pc^3}$ per yr. HAN and ITY denote  \cite{ity97} and \cite{han98} models, OBS - observational data. 
\end{tablenotes}
\label{tab:mod}
\end{table}

Briefly, the comparison of models shows the following.
Models with an exponential SFR compared to the models with a constant SFR (mod.\,A vs. mod.\,C) have  a higher number of old stars and a higher mass of the disk. This gives higher birthrates of
CDWD and AM\,CVn stars. The rate of mergers giving SNe\,Ia is similar, since it is determined  mainly by the SFR in the past $\sim 300$\,Myr\ \cite{ty94}. 

Model A with  Eq.~(\ref{eq:aml}) for the first mass loss episode 
gives less mergers than model $A^\prime$ with Eq.~(\ref{eq:ce}), since it produces  wider pairs. For the same reason the occurrence of SNe\,Ia is lower in model $A^\prime$.
Formation rate of interacting systems (AM CVn's) is higher in model $A^\prime$, because, if two common envelope phases occur, the second-formed WD is typically less massive than companion; if such a pair is brought into contact due to angular momentum loss via gravitational wave radiation, unequal masses of components  favour stable mass transfer \cite{npvy01}.  

Model D has an IMF rather similar to Han's model, but treats the first mass loss episode differently , does not have companion reinforced stellar wind, and has a higher $\alpha\lambda$ value. This results in relatively less mergers in the first stage of mass loss and in higher birthrate and total number of CDWD and higher occurrence rate of  SNe\,Ia.

The Iben et al. model differs from all other models by  applying   a different equation for the evolution in the common envelope, which is, in practice, equivalent to the usage of much higher value of $\alpha\lambda$. This results in less frequent mergers in both stages   of mass loss, hence, a higher total number of CDWD. The total number of the Galactic WD in this model is lower, compared to our models. This is due to the ITY assumption that close and wide binaries obey different distributions over the mass ratios of the components: flat for close systems and $\propto\!q^{-2.5}$\ for wide systems (see Tutukov \& Yungelson (1993)\nocite{ty93} for details).   

Our models share with the Han and ITY models the same assumptions on the initial distributions of close binaries over mass ratios of components and their separations and have rather similar initial IMF for the primary components.  The variations of the birthrates  and numbers of objects in different classes (within factor $\sim 3$) arise mainly from the differences  in the treatment of mass loss and transfer, in initial-final mass relations and other more minor details of population synthesis codes. Since the assumptions in the different studies are generally in agreement with results obtained from the modelling of stellar evolution,
Tab.~\ref{tab:mod} illustrates the limits of the accuracy of predictions  by the state of the art population synthesis studies for binary stars.    

Observational data which may help to constrain the models are rather scarce and uncertain, e.g., the estimates of  local space density of white dwarfs $\rho_{wd,\odot}$ differ by a factor 5 \cite{khh99,fes98,osw+95,rt95}. Our basic model A, as well as model C, complies with  these  observational limits.  Model A gives the annual birthrate of planetary nebulae in somewhat better agreement with the observations \cite{pot96} than model C. 

Model A, as well as the rest of the models in Tab.~\ref{tab:mod}, agrees reasonably  with the occurrence of SNe\,Ia in galaxies similar to the Milky Way (E. Cappellaro, this volume). 
The difference in SNe\,Ia rates between the models has to be attributed mainly  to the different treatment of mass loss and to the different initial-final mass relations for the components of binaries.  

\section{Models {\it vs.} observations}
\label{sec:mod_vs_obs}

\begin{figure*}[!t]
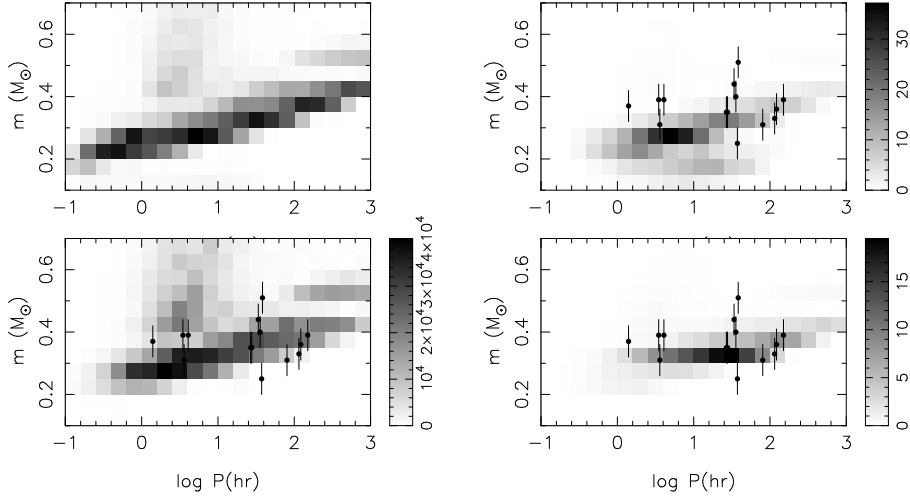

\begin{minipage}{0.45\textwidth}
\includegraphics[angle=-90,scale=0.23]{ynpzv_f4_tl.ps}
\end{minipage}
\hspace*{0.055\textwidth}
\begin{minipage}{0.45\textwidth}
\includegraphics[angle=-90,scale=0.23]{ynpzv_f4_tr.ps}
\end{minipage}
\\
\begin{minipage}{0.45\textwidth}
\vspace*{-1cm}
\includegraphics[angle=-90,scale=0.23]{ynpzv_f4_bl.ps}
\end{minipage}
\hspace*{0.055\textwidth}
\begin{minipage}{0.45\textwidth}
\vspace*{-1cm}
\includegraphics[angle=-90,scale=0.23]{ynpzv_f4_br.ps}
\end{minipage}
\caption[]{The model population of CDWD as a function of the orbital
  period and mass of the brighter dwarf of the pair. Top left:  distribution of the currently born CDWD in model A. Top right: ``observed sample'' ($V_{\rm lim} = 15)$, with cooling according to DSBH and Bl\"ocker (1995)\nocite{blo95b}. Bottom right: the same sample for the modified DSBH cooling.
Bottom left: the total Galactic population of CDWD younger than 100\,Myr. Dots with error bars: observed CDWD. }
\label{fig:P_m}
\end{figure*}

{\textit{\textbf{1. Orbital periods and masses of close double white dwarfs.}}}
The  parameters which are determined for all known CDWD are the orbital period  and the mass of the brighter
component of the pair. We plot in Fig.~\ref{fig:P_m} the
$P_{\rm orb} - m$\ distributions of  the occurrence rate for the currently  born CDWD and for the simulated
magnitude limited samples ($V_{\rm lim}$ = 15) for the models with different cooling prescriptions.

\begin{figure}[!ht]
\begin{tabular}{lr}
\begin{minipage}[!ht]{0.45\textwidth}
\includegraphics[angle=-90,scale=0.255]{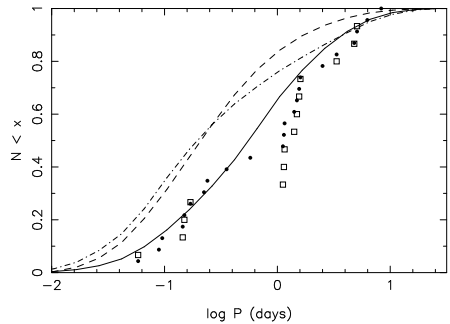}
\end{minipage}
\hspace{0.05\textwidth}
\begin{minipage}[!t]{0.45\textwidth}
\vspace{-0.9cm}
\caption{Cumulative distribution over the periods. The solid line is for the model with reduced cooling time for low-mass WD. The dashed line is for DSBH cooling without
  modifications and the dash-dotted line is for a model with constant ``observability''
  time of 100\,Myr.  Open squares give distribution of observed CDWD, filled circles give the distribution including the  sdB+wd  binaries. }
\end{minipage}
\end{tabular}
\label{fig:Pcumul}
\end{figure}

  In general, the  white dwarf observed as the brighter member of the pair, is the one that was formed last, but occasionally, it may be the one that was formed first, due to the effect of differential cooling, see Fig.~\ref{fig:wdcool}.
The top right panel of Fig.~\ref{fig:P_m} shows that if DSBH cooling curves are taken at face value, the observed sample contains predominantly low mass ($M\!\aplt\!0.3\,\ms$) white dwarfs, in contradiction to observations. However, as we mentioned above, low-mass WD may experience thermal flashes, which may reduce the mass of their hydrogen envelopes and extinguish hydrogen burning. This may be especially true for WD in close pairs $(a\!\sim\!\rs)$, where one  easily expects the formation of a common envelope during a flash. Since  estimates for the  amount of mass which may be lost in a flash is not yet available , we make an {\em ad hoc} extreme assumption that white dwarfs with masses below 0.3\,\ms\ cool like the most massive helium (0.46\,\ms) white dwarfs (hereafter we call this 
{\textit{modified DSBH cooling}}). As can be seen from the bottom right panel of Fig.~\ref{fig:P_m}, this assumption brings the model in a much better agreement with observations. All model distributions which follow, are given for  the modified DSBH cooling. 

For comparison, we plot in the left bottom panel of   Fig.~\ref{fig:P_m} the ``observed'' distribution assuming (like in the studies of other authors)  that all WD are visible for $10^8$\,yr, unless a pair merges earlier due to GWR. Since in this case  cooling curves are not applied,  i.e. no magnitudes are
computed for the white dwarfs, we can not construct a magnitude limited
sample and this panel gives the total number of ``potentially visible'' CDWD in the Galaxy. The better agreement with observations  compared to the modified DSBH case is only apparent, as can be easily seen from the cumulative distributions (Fig.~\ref{fig:Pcumul}).

Figure~\ref{fig:Pcumul} shows a deficiency of observed systems between $\sim\!5$\,hr and $\sim\!1$\,day.  No selection effects are known that prevent detection of white dwarf binaries with such periods. This ``gap'' may be partially filled if we plot also subdwarf B stars with suspected white dwarf companions, thus assuming that the current sdB star is a white dwarf in the making.  In addition to systems listed in Tab.~\ref{tab:obs}, we include  the binaries  for which only $P_{orb}$ is determined. However, the number of detected systems is still too small to decide whether the ``gap'' is real and whether revisions of the stellar evolution models  or CDWD formation scenarios are required.        

The merger of  white dwarfs is one of the models for SNe\,Ia (see, e.g., Livio 1999\nocite{liv99} for a review).  We
estimate that one merger candidate with $M_1+M_2 \geq 1.4\,\ms$\ may be found in a WD sample complete to $V \approx 15$ which contains  $\sim 200$ CDWD among a total of  $\sim 1000$  WD.  In  view of the data in  Tab.~\ref{tab:mod}, this estimate is probably uncertain within a factor of  at least $2\div3$. 

\begin{figure}[!t]
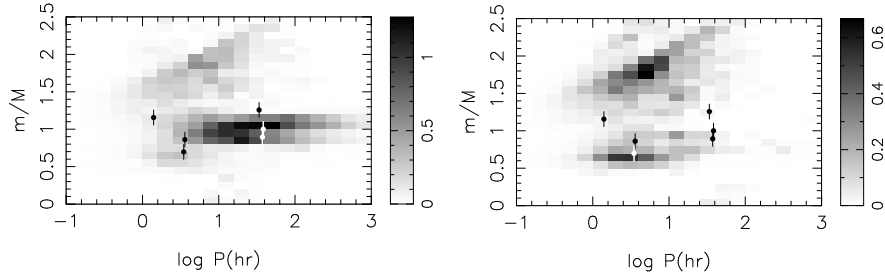

\begin{tabular}{lr}
\begin{parbox}{0.48\textwidth}
{
\includegraphics[angle=-90,scale=0.23]{ynpzv_f6_l.ps}
\hspace{0.48\textwidth} 
}
\end{parbox}
\begin{parbox}{0.48\textwidth}
{
\includegraphics[angle=-90,scale=0.23]{ynpzv_f6_r.ps}
}
\end{parbox}
\end{tabular}
\caption[]{Model sample of  CDWD as a function of the orbital
period and mass ratio of components for $V_{lim}=15$ and
the ratio of the luminosities of the components $\leq 5$. Left panel is for model A with   the  first phase of mass transfer treated by Eq.~(\ref{eq:aml}), right panel is the same for the model A$^\prime$\ with standard  Eq.~(\ref{eq:ce}) with $\alpha\lambda=2$\ in both mass transfer phases. Dots with error bars  are observed systems.}
\label{fig:P_q}
\end{figure}

{\textit{\textbf{2. Period - mass ratio distribution.}}}
Our treatment of the first
phase of mass transfer between a giant and a main-sequence star, which doesn't result in a significant spiral-in, leads to  a  concentration of the mass ratios of the model systems around  
$q\!\sim\!1.$ In
practice,  $q$  in the observed systems can only
be determined if the ratio of the luminosities  of the components is $ \aplt 5$ \cite{mmm00}. The $P_{orb}-q$ distribution for the theoretical magnitude limited sample which obeys this criterion, is  shown in Fig.~\ref{fig:P_q}. 
In model A$^\prime$ (with a common envelope in the first mass transfer) CDWD have predominantly  $q\!\sim\!(1.75 - 2)$, which is not observed and  there are hardly any systems with $q\!\sim\!1$. 

\begin{figure}[!t]
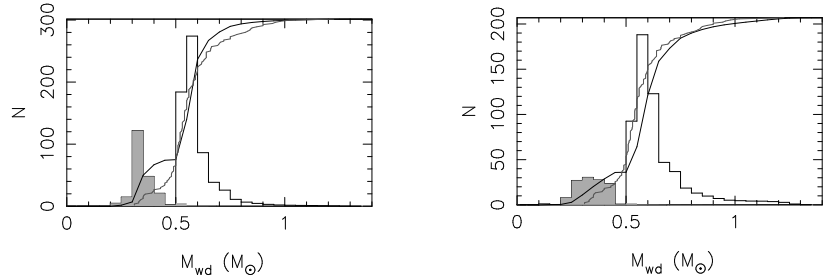

\begin{tabular}{lr}
\begin{parbox}{0.48\textwidth}
{
\includegraphics[angle=-90,scale=0.23]{ynpzv_f7_l.ps}
\hspace{0.48\textwidth} 
}
\end{parbox}
\begin{parbox}{0.48\textwidth}
{
\includegraphics[angle=-90,scale=0.23]{ynpzv_f7_r.ps}
}
\end{parbox}
\end{tabular}
\caption[]{Left panel - mass spectrum of all white dwarfs in model A with modified DSBH cooling. Right panel - the same for a model in which all helium white dwarfs
  cool like a $\sim 0.414\,\msun$\ dwarf and all CO white dwarfs cool like a
  0.605\,\msun\ dwarf. In both panels the theoretical cumulative distribution is shown as a solid black line  and the
cumulative distribution of observed systems as a grey line. }
\label{fig:mass}
\end{figure}

{\textit{\textbf{3. Mass spectrum of white dwarfs.}}}
The left panel of Fig.~\ref{fig:mass} shows the  
spectrum of white dwarf masses in a sample limited by $V =
15$ and based on model A. It  includes white dwarfs in close pairs which are brighter than their
companions and genuine single  objects, white dwarfs which are components of the initially wide pairs, merger products, white
dwarfs which became single due to disruption of binaries  by SNe
explosions. In the same Figure we plot the cumulative distribution for DA white dwarfs masses, as estimated by Bergeron et al. (1992) \nocite{bsl92} and Bragaglia et al. (1995)\nocite{brb95}\footnote{These masses may have to be
increased by about 0.05\,\msun, if one uses models of
white dwarfs with thick hydrogen envelopes for the  estimates 
\cite{ngs99}.}.

The  expected fraction of CDWD among all ``observed'' WD in our preferred model A is $\sim 23\%$, slightly higher than the upper limit of the range suggested by Maxted and Marsh (1999)\nocite{mm99} for DA white dwarfs:
1.7 to 19\% with 95\% confidence. Since lowering the initial fraction of binaries  below 50\% would contradict the observations, the high percentage of CDWD in the model sample may mean that, even after our  modification,  DSBH results overestimate cooling times of the lowest mass white dwarfs. The right panel of Fig.~\ref{fig:mass} shows results of a simple numerical experiment in which we assign to all helium white dwarfs the cooling curve of  a 0.414\,\msun\ dwarf
 from DSBH and a cooling curve of a
0.605\,\msun\ dwarf \cite{blo95b} to all CO white dwarfs. This gives 17\% for
 the fraction of CDWD. Even so, the contribution  of the lowest mass WD to the total model sample still seems to be overestimated. However, there may exist yet unknown selection effects which prevent their detection.  

Yet another problem is the deficiency of $(0.45 - 0.50)\,\ms$\ white dwarfs in the model sample.
In this part of the theoretical mass distribution  only hybrid white dwarfs, descending via  case B mass transfer from the stars in a narrow range of $\sim\!(2.5 - 5)\,\ms$ do occur \cite{it85}. Their formation rate is relatively low. On the other hand, cooling of these objects which have almost pure helium envelopes with mass  $\sim\!(0.01 - 0.1)$\,\ms\ was not yet studied and it's possible that the simple interpolation between cooling curves for CO and He white dwarfs is not appropriate and gives misleading results.  

\section{Conclusions}

1. The standard Algol-type evolution and spiral-in in common envelopes  cannot explain the wide orbits of the progenitors of close double WD  after the first stage of mass loss.
In  low-mass binaries with similar masses of components,  the loss of the envelope of the
initial primary probably doesn't cause a strong spiral-in. 
In the absence of a physical picture of this process, we suggest to describe the change of separation of the components by an angular momentum loss law [Eq.~(\ref{eq:aml})]. An attempt to model the parameters of three observed CDWD with two helium components results in $1.4 \aplt \gamma \aplt 1.8$. 

2.  Modelling of the observed sample of close binary WD shows that low-mass WD have to cool faster than is suggested by the  
recently published cooling curves for helium WD with thick hydrogen envelopes \cite{dsb+98,sea00}. 
We suggest that an additional mass loss by WDs in winds or thermal flashes may strongly diminish the masses of envelopes and reduce the cooling times.     

3. A reasonable agreement of the population synthesis results with observations in  the estimates of the local space density of WD, and the  $P_{orb}- m$, $P_{orb} -q$, and $N(P_{orb})$ distributions for  CDWD may be achieved if the parameter $\gamma$ in Eq.~(\ref{eq:aml})  is  $\sim\!1.7$ and it's assumed that WD with 
$M\!\leq\!0.3$\,\ms\ cool like the most massive helium WD.  Further reduction of cooling times for the helium WD may even improve agreement. 

4. The model with  an initial binary fraction of 
  50\%\ (2/3 of stars are in binaries) fits the observations better than  the model in which  initially 100\% of stars are in  binaries.

5. We estimate that the detection  of at least one WD pair  with $M_1+M_2 \geq M_{Ch}$ may require  a survey of $\sim 1000$ white dwarfs. But  the uncertainties in the input data for the population synthesis studies and badly known selection effects, most probably make this estimate accurate only within a factor of $\sim 2\div 3$.
   
\vspace{0.3cm} 
LRY  acknowledges the support of the LOC and 
Astronomical Institute ``Anton Pannekoek'' which enabled him to
attend the conference.  This work was supported by NWO Spinoza grant
08-0 to E.P.J.~van den Heuvel, RFBR grant 99-02-16037, 
``Astronomy and Space Research'' Program (project 1.4.4.1), and NASA through
Hubble Fellowship grant HF-01112.01-98A awarded (to SPZ) by the Space
Telescope Science Institute, which is operated by the Association of
Universities for Research in Astronomy, Inc., for NASA under contract
NAS\, 5-26555.

\bibliographystyle{apalike}
\chapbblname{ynpzv}
\chapbibliography{}

\end{document}